\documentclass{article}

\usepackage{gensymb}
\usepackage{PRIMEarxiv}

\usepackage[utf8]{inputenc} 
\usepackage[T1]{fontenc}    
\usepackage{hyperref}       
\usepackage{url}            
\usepackage{booktabs}       
\usepackage{amsfonts}       
\usepackage{nicefrac}       
\usepackage{microtype}      
\usepackage{lipsum}
\usepackage{fancyhdr}       
\usepackage{graphicx}       
\graphicspath{{media/}}     

\title{Highly Oriented PZT Platform for Polarization-Independent Photonic Integrated Circuit and Enhanced Efficiency Electro-Optic Modulation}

\author{
  Suraj, Shankar Kumar Selvaraja \\
  Center for Nanoscience and Engineering \\
  Indian Institute of Science, Bangalore 560012, India \\
  \texttt{\{Suraj,Shankar Kumar Selvaraja\} suraj2@iisc.ac.in, shankarks@iisc.ac.in} \\
}

\begin{document}
\maketitle

\vspace{-2em}
\begin{abstract}
We demonstrate, for the first time, sputtered PZT as a platform for the development of Si-based photonic devices such as rings, MZI, and electro-optic modulators. We report the optimization of PZT on MgO(002) substrate to obtain highly oriented PZT film oriented towards the (100) plane with a surface roughness of 2 nm. Si gratings were simulated for TE and TM mode with an efficiency of -2.2 dB/coupler -3 dB/coupler respectively with a polarization insensitive efficiency of $\approx$50\% for both TE and TM mode. Si grating with an efficiency of around -10 dB/coupler and a 6 dB bandwidth of 30 nm was fabricated. DC Electro-optic characterization for MZI yielded a spectrum shift of 71 pm/V at the c-band.
\end{abstract}


\section{Introduction}
\vspace{-0.6em}
PZT in recent times has become a material of choice in integration with Si to fabricate non-linear photonic devices. PZT has two important properties that make it desirable for photonics. They are a high bandgap(~4.1eV) \cite{du1998crystal}  and a large non-linear Pockels coefficient \cite{feutmba2020strong,spirin1998measurement}. Due to these advantages, PZT is now being used to perform Electro-optic modulation by integrating with Si substrate or SiN substrate \cite{alexander2018nanophotonic,singh2021sputter,zouboulis1991refractive}. Most of the reports on PZT modulators have performance limitations due to the poly-crystallinity of PZT and due to the limited interaction of PZT with the optical field. In our previous work\cite{singh2021sputter}we had performed modulation using sputtered deposited PZT on MgO buffer. MgO was chosen due to low refractive index \cite{zouboulis1991refractive} and a small lattice constant difference between PZT and MgO\cite{gilmore2003growth,tonejc2002analysis,yano1994epitaxial}.In this work, we have tried to overcome both bottlenecks by starting with optimizing the PZT on MgO (002) substrate. The figure of merit used to qualify the PZT film are phase, surface roughness, and the presence of ferroelectricity. PZT was optimized to obtain a morphotropic phase boundary which yields maximum piezo-electric coefficient\cite{hiboux1999domain,karapuzha2016structure}. Losses in a photonic device are very susceptible to surface roughness. Oriented PZT with very small surface roughness, photonic device fabrication using PZT as a platform becomes feasible. By using PZT itself as a platform rather than using it as a material to integrate on top of Si waveguide, the optical field can interact directly with PZT and hence potentially increase the efficiency of the electro-optic modulator such as DC spectrum shift as well as reduce V$_\pi$L$_\pi$ values leading to low voltage and power operation. In this work, we have demonstrated a tunable Si grating for TE, TM photonic device on PZT platform with an efficiency of 60\%, 50\% respectively. A polarization-independent operation was simulated with an efficiency of 50\% for TE and TM mode. Electro-optic modulation was demonstrated on MZI with a DC shift of 71 pm/V.

\section{Experimental}
\subsection{PZT film growth and characterization}
\begin{figure}[htbp]
\centering\includegraphics[width=1.05\linewidth]{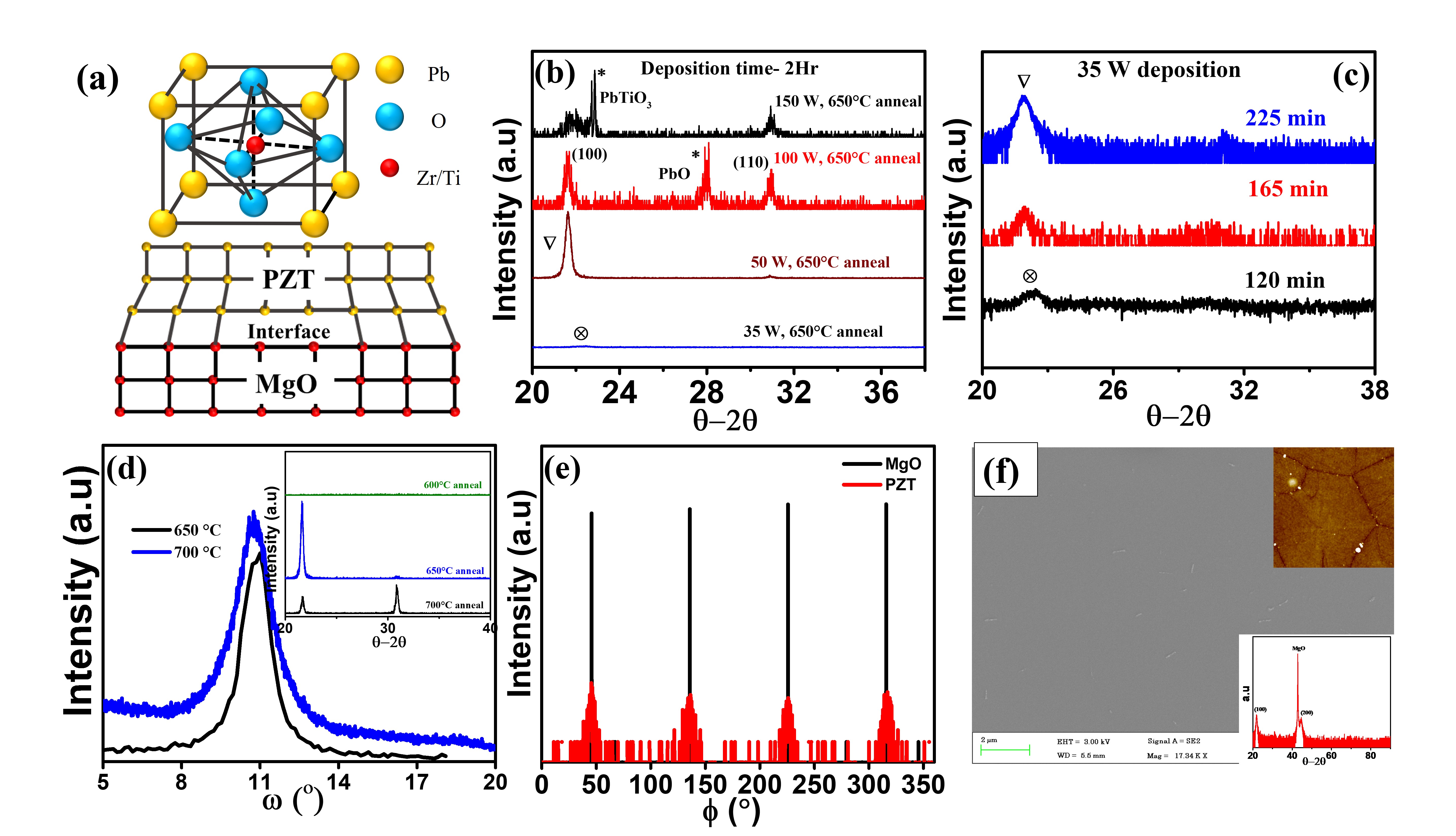}
\caption{(a) Lattice structure of PZT and PZT/MgO stack; (b) HRXRD showing a normal scan of the deposited PZT at varying RF power and annealed at 650\degree C; (c) HRXRD showing a normal scan of the deposited PZT 35 W RF power with varying deposition time and annealed at 650\degree C; (d) $\omega$ scan of (100) peak for the PZT annealed at 650 and 700\degree C  with inset showing HRXRD normal scan at different annealing temperature; (e) $\phi$ scan of 50 W, 650\degree annealed PZT sample; (f) SEM, AFM and HRXRD of the optimized PZT film.}
\label{fig:material}
\end{figure}

Fig.\ref{fig:material}(a) shows the lattice of PZT with the "Pb" atom occupying the vertices, "O" at the face centers and "Zr" and "Ti" at the body centers slightly displaced from the center giving it a spontaneous polarization. Fig.\ref{fig:material}(a) shows the PZT/MgO stack wherein MgO with a lattice constant of (a=0.4216 nm) puts the PZT(a=0.404 nm) film under tensile strain. The quality of the deposited and annealed PZT film is dictated by the stress developed in the deposited film. As reported in our work\cite{singh2021sputter}, we use a small ramp rate to mitigate the sudden stress development during perovskite phase formation. We use RF sputter deposited PZT and ex-situ annealing to form perovskite phase film on MgO(002) substrate. The PZT deposition was done at varying RF power of 35 W,50 W, 100 W, and 150 with a source-to-target distance of 4.2 cm and 14 sccm argon flow. The deposition was done for 2 Hr followed by ex-situ anneal in air ambiance with a ramp rate of 1.5 \degree C/min and dwell temperature of 650\degree C. The deposited sample was allowed to cool to room temperature over a duration of 10 Hr. Fig.\ref{fig:material}(b) shows the high-resolution X-ray diffraction (HRXRD) spectra of the PZT phase obtained for varying RF power (deposition time of 2 Hr) showing a polycrystalline PZT film on MgO(002) at 150W, highly oriented PZT  with only [100] plane for the film deposited at 50 W and at 35 W showing the presence of PbO and PbTiO$_3$ ("$\otimes$"). Fig.\ref{fig:material}(c) shows the variation in the phase of PZT deposited at RF power of 35 W, 650\degree C annealing temperature, and varying deposition time with perovskite phase($\nabla$) appearing for deposition of $\geq$165 min confirming the thickness dependency of perovskite phase formation at a constant annealing temperature. PbO and PbTiO$_3$ are formed at the interface which on further annealing forms the perovskite phase\cite{chen1996texture,chen1998texture} but also leads to an increase in the crack density of the film. The degree of crystallinity can be increased by increasing the annealing temperature as seen in ($\omega$ scan) in Fig.\ref{fig:material}(d) with the reduction in FWHM with increasing temperature. We observe a reduction in the FWHM at 700\degree C film, but it increases the roughness and also other peaks start to apper in the XRD, thus reducing crystallinity as seen in the inset in Fig.\ref{fig:material}(d). The periodic peaks of PZT in in-plane XRD spectra($\phi$ scan) shown in Fig.\ref{fig:material}(d), deposited at 50 W and 650\degree C annealing temperature, further confirms the high orientation of the PZT film on MgO for 50 W deposited sample. FESEM image of the deposited and annealed PZT film in Fig.\ref{fig:material}(f) shows a crack-free film with AFM measurements giving a roughness of less than 2 nm with the inset showing the XRD spectra and AFM image of the highly oriented PZT film. 

\section{Design and Simulation}
\subsection{Waveguide design}

\begin{figure}[t]
\centering\includegraphics[width=\linewidth]{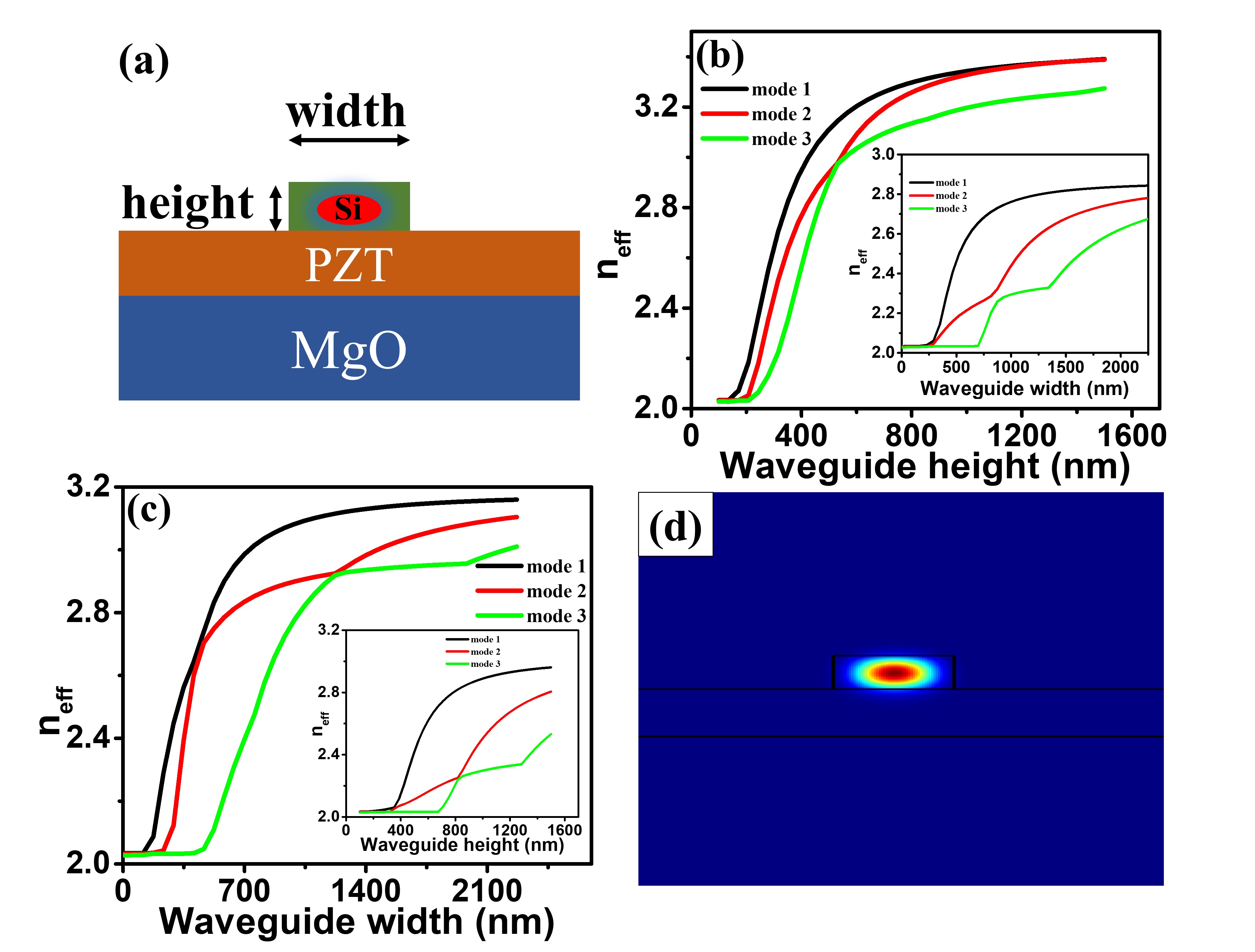}
\caption{(a) The schematic of the proposed Si on PZT waveguide design. Waveguide dispersion with variation in the waveguide; (b) height (with waveguide width = 1.3 $\mu$m); (c) waveguide width (at height = 411 nm); (d) mode field confinement at 1.3 $\mu$m waveguide width and 411 nm waveguide height.
}
\label{fig:mode_sim}
\end{figure}

Fig.\ref{fig:mode_sim} shows the cross-section schematic of Si waveguide on PZT/MgO platform used to perform mode simulation with the optical field confined predominantly confined in the Si waveguide. Fig.\ref{fig:mode_sim}(b) gives the waveguide dispersion curve with the single moded operation achieved at a waveguide dimension of $\approx$466 nm width and $\approx$244 nm height with TE as the fundamental mode. Fig.\ref{fig:mode_sim}(c) on the other hand gives a TM fundamental operation with a dimension of $\approx$290 nm width and $\approx$495 nm height.Fig.\ref{fig:mode_sim}(d) gives the fundamental TE mode for waveguide dimension of 1.3 $\mu$m width and 411 nm height. 

\subsection{Grating coupler design simulation}

\begin{figure}[htbp]
\centering\includegraphics[width=0.8\linewidth]{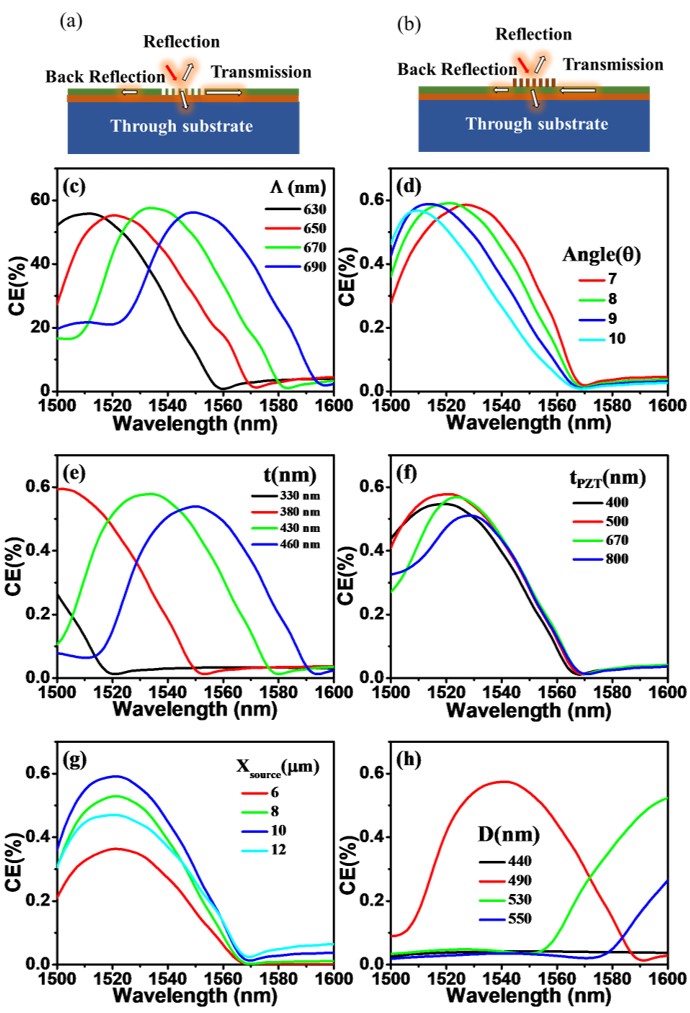}
\caption{Cross-section schematic for simulation of (a) in-coupling and (b) out-coupling; Variation of Si grating coupler CE with (c) $\Lambda$(t=411 nm,t$_{PZT}$=534 nm,$\theta$= 8\degree,X$_{source}$=10$\mu$m, D = 477 nm), (d) $\theta$($\Lambda$=650 nm,t$_{PZT}$=534 nm,t=411 nm,X$_{source}$=10$\mu$m, D = 477 nm),(e) t($\Lambda$=650 nm,t$_{PZT}$=534 nm,$\theta$= 8\degree,X$_{source}$=10$\mu$m, D = 477 nm),(f) t$_{PZT}$($\Lambda$=650 nm,X$_{source}$=10$\mu$m,t=411 nm,$\theta$= 8\degree, D = 477 nm), (g) source position($\Lambda$=650 nm,t$_{PZT}$=534 nm,t=411 nm,$\theta$= 8\degree, D = 477 nm),
(h) D($\Lambda$=650 nm,X$_{source}$=10$\mu$m,t=411 nm,$\theta$= 8\degree, t$_{PZT}$=534 nm),for TE mode.}
\label{fig:TE}
\end{figure}

\begin{figure}[htbp]
\centering\includegraphics[width=\linewidth]{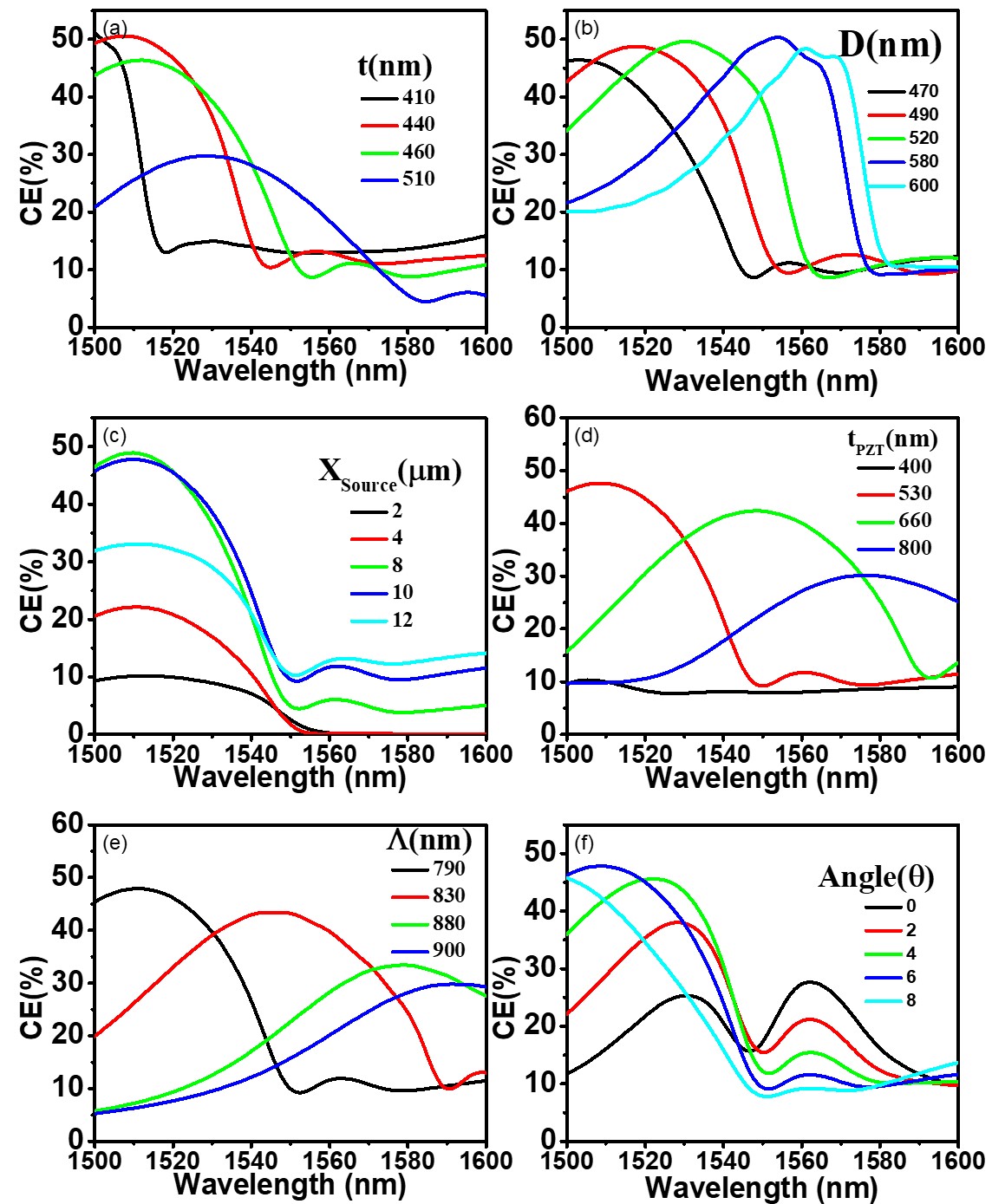}
\caption{Variation of Si grating coupler CE with (a) t($\Lambda$=789 nm,t$_{PZT}$=534 nm,$\theta$= 5.8\degree,X$_{source}$=10$\mu$m, D = 477 nm),(b) D($\Lambda$=789 nm,X$_{source}$=10$\mu$m,t=454 nm,$\theta$= 5.8\degree, t$_{PZT}$=534 nm),(c) source position($\Lambda$=789 nm,t$_{PZT}$=534 nm,t=454 nm,$\theta$= 5.8\degree, D = 477 nm),(d) t$_{PZT}$($\Lambda$=789 nm,X$_{source}$=10$\mu$m,t=454 nm,$\theta$= 5.8\degree, D = 477 nm),(e) $\Lambda$(t=454 nm,t$_{PZT}$=534 nm,$\theta$= 5.8\degree,X$_{source}$=10$\mu$m, D = 477 nm),(f) $\theta$($\Lambda$=789 nm,t$_{PZT}$=534 nm,t=454 nm,X$_{source}$=10$\mu$m, D = 477 nm),
for TM mode.}
\label{fig:TM}
\end{figure}

\begin{figure}[htbp]
\centering\includegraphics[width=\linewidth]{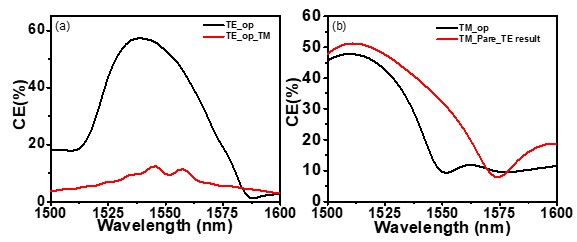}
\caption{Simulated transmission characteristic of Si grating coupler demonstrating (a) polarization dependency at optimized TE parameter($\theta$=8\degree, $\Lambda$=645 nm,t$_{PZT}$=534 nm,t=411 nm, X$_{source}$=10$\mu$m, D = 477 nm)) and (b) polarization independent characteristic ($\theta$=5.8\degree, $\Lambda$=789 nm,t$_{PZT}$=534 nm,t=454 nm, X$_{source}$=10$\mu$m, D = 477 nm)).
}
\label{fig:Polarization_independent}
\end{figure}


The waveguide dispersion optimization shows that there is a small difference in the effective refractive index of TE and TM mode making a polarization-independent grating coupler possible. We had shown a polarization-independent Si grating coupler on GaN[\textbf{reference}] with a refractive index of $\approx$2.3 which is similar to PZT. Fig.\ref{fig:TE}(a) and (b) show the cross-section schematic of the Si grating coupler used to in-couple and out-couple the Gaussian TE and TM source into and out of the waveguiding medium. Fig.\ref{fig:TE}(c-h) and Fig.\ref{fig:TM}(a-f) show the optimization of the Si grating with varying parameters such as grating period ($\Lambda$), Gaussian source angle ($\theta$), grating thickness (t), PZT thickness (t${_PZT}$), source position (X${_source}$) and duty cycle (D) for TE and TM mode respectively. A peak coupling efficiency (CE) of 60\% was obtained for TE mode with an optimized "$\Lambda$" of 646 nm, a "t" of 411 nm, "D" of 477 nm, "t${_PZT}$" of 534 nm, "$\theta$" of 8\degree and at "X${_source}$" of 10 $\mu$m. The corresponding optimized parameter for TM Gaussian source for a peak coupling of 50\% is "$\Lambda$" of 789 nm, a "t" of 454 nm, "D" of 477 nm, "t${_PZT}$" of 534 nm, "$\theta$" of 5.8\degree and at "X${_source}$" of 10 $\mu$m. Fig.\ref{fig:Polarization_independent}(a) shows the polarization selective characteristic wherein Si grating optimized for TE mode is highly selective with an $\approx$60\% CE for TE mode and $\approx$10\% for a TM mode while Si grating optimized for TM mode shows a polarization insensitive characteristic with $\approx$50\% CE for both TE and TM mode. Table.\ref{tab:sensitivity} shows that both TE gratings are more sensitive to the $X{_{source}}$ and span while TM gratings are more sensitive to $X{_{source}}$ and $\theta$ with both TE and TM being relatively insensitive to small variation in other parameters. Table.\ref{tab:Losses} confirms the substrate leakage as a major source of loss and can be reduced by employing a Bragg reflector in the proposed waveguide architecture as Si/PZT/Bragg/MgO.

\begin{table}[]
\centering
\caption{Sensitivity of coupling to various parameters}
\label{tab:sensitivity}
\begin{tabular}
{|c|c|c|c|c|c|}
\hline
Parameter & Sensitivity & &Parameter & Sensitivity &  \\
\hline
 & TE & TM & & TE & TM \\
\hline
$\theta$     & 0.38\%/$\theta$(\degree) & 1.1\%/$\theta$(\degree) &$t{_{PZT}}$     & 0.0051\%/nm & 0.04\%/nm  \\
\hline
t     & 0.0308\%/nm & 0.023\%/nm  &$X{_{source}}$     & 3\%/$\mu$m & 3.9\%/$\mu$m      \\
\hline
$\Lambda$     & 0.0695\%/nm & 0.11\%/nm  &D     & 1\%/nm & 0.012\%/nm  \\
\hline
\end{tabular}
\end{table}

\begin{table}[]
\centering
\caption{Otical power distribution}
\label{tab:Losses}
\begin{tabular}
{|c|c|c|c|c|c|}
\hline
Parameter & TE & TM & Parameter & TE & TM   \\
\hline
Through grating     & 2\% & 15\% &Substrate loss     & 20\% & 35\%  \\
\hline
back reflection     & 18\% & 5\% &Coupling     & 60\% & 45\%  \\
\hline
\end{tabular}
\end{table}

\section{Fabrication}

\begin{figure}[htbp]
\centering\includegraphics[width=\linewidth]{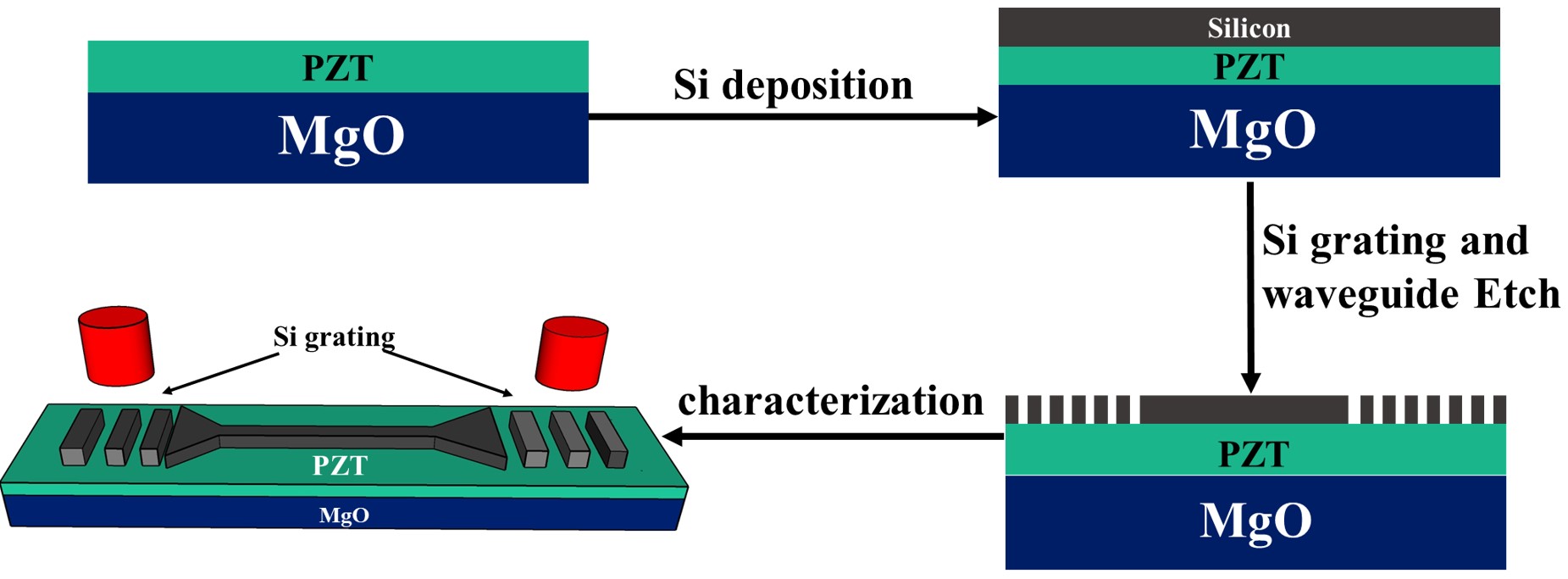}
\caption{Schematic for device fabrication process flow.}
\label{fig:process}
\end{figure}

Fig. 6 shows the process flow to fabricate the E-O modulator. The fabrication starts with the deposition of oriented PZT on MgO. An added step of depositing a buffer layer of Al$_2$O$_3$ (10 nm) on top of PZT was done. The purpose of adding the buffer layer is to make the process CMOS compatible as Pb can contaminate the fabrication line. After this amorphous Si of thickness around 395 nm was deposited followed by deposition of 16 nm of doped amorphous Si. Doped Si layer helps in fabrication process as it reduces the effect of charging when doing e-line lithography. A single step of 411 nm is etched to obtain a Si waveguide on PZT on MgO platform. After passive device characterization, electrodes are deposited around the MZI and ring resonator to form E-O device.

\section{Measurements and Discussion}
\subsection{Passive device characterization}
\begin{figure}[htbp]
\centering\includegraphics[width=\linewidth]{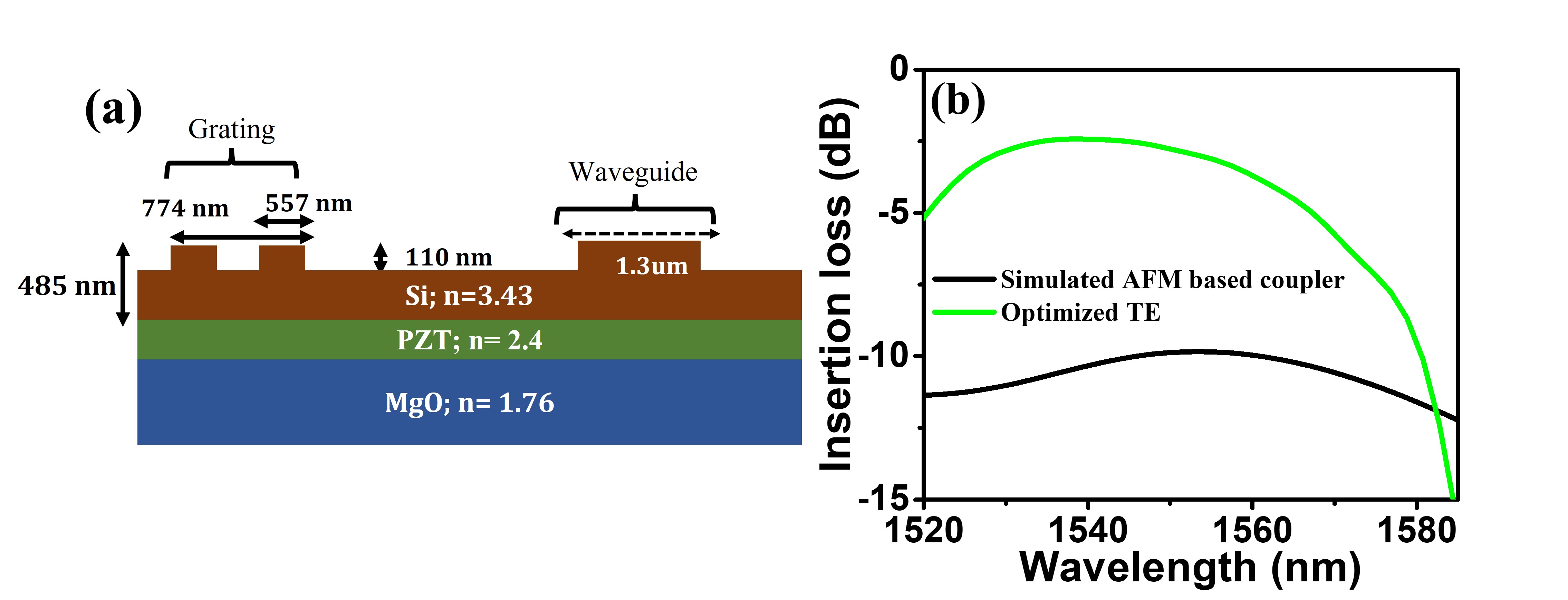}
\caption{(a) Schematic of the fabricated device showing dimension measured from AFM scanning; (b) comparison of the CE for the characterized grating couplers and simulated grating coupler.}
\label{fig:device_char}
\end{figure}

\begin{figure}[t]
\centering\includegraphics[width=\linewidth]{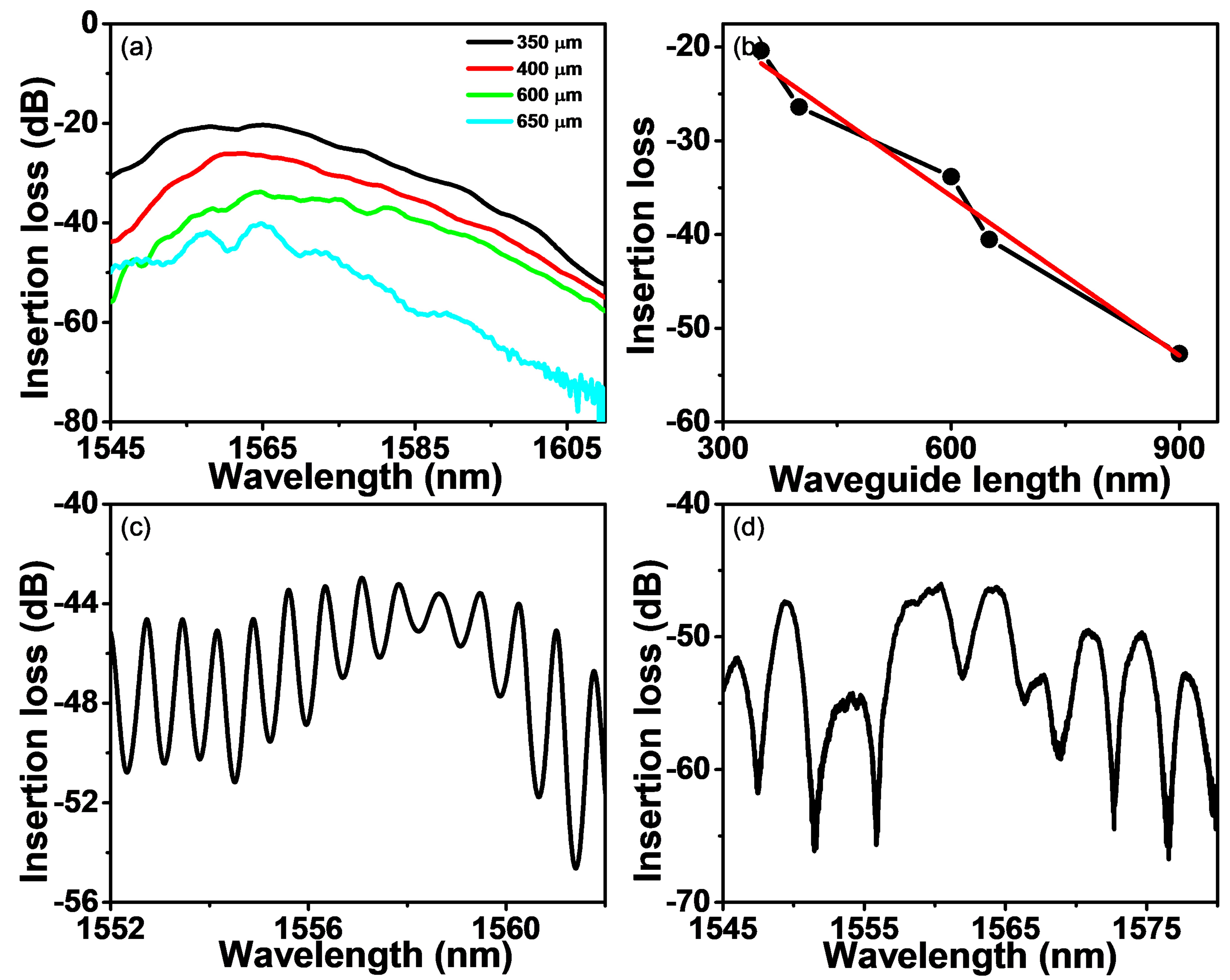}
\caption{(a)Measured optical insertion loss for the waveguide of varying length with $\Lambda$ = 774 nm, D = 557 nm, t= 110 nm, t$_{PZT}$ = 650 nm; (b) Waveguide loss calculation using insertion loss vs waveguide length slope; (c) fabricated ring response and (d) fabricated MZI response. }
\label{fig:Passive}
\end{figure}

\begin{figure}[t]
\centering\includegraphics[width=\linewidth]{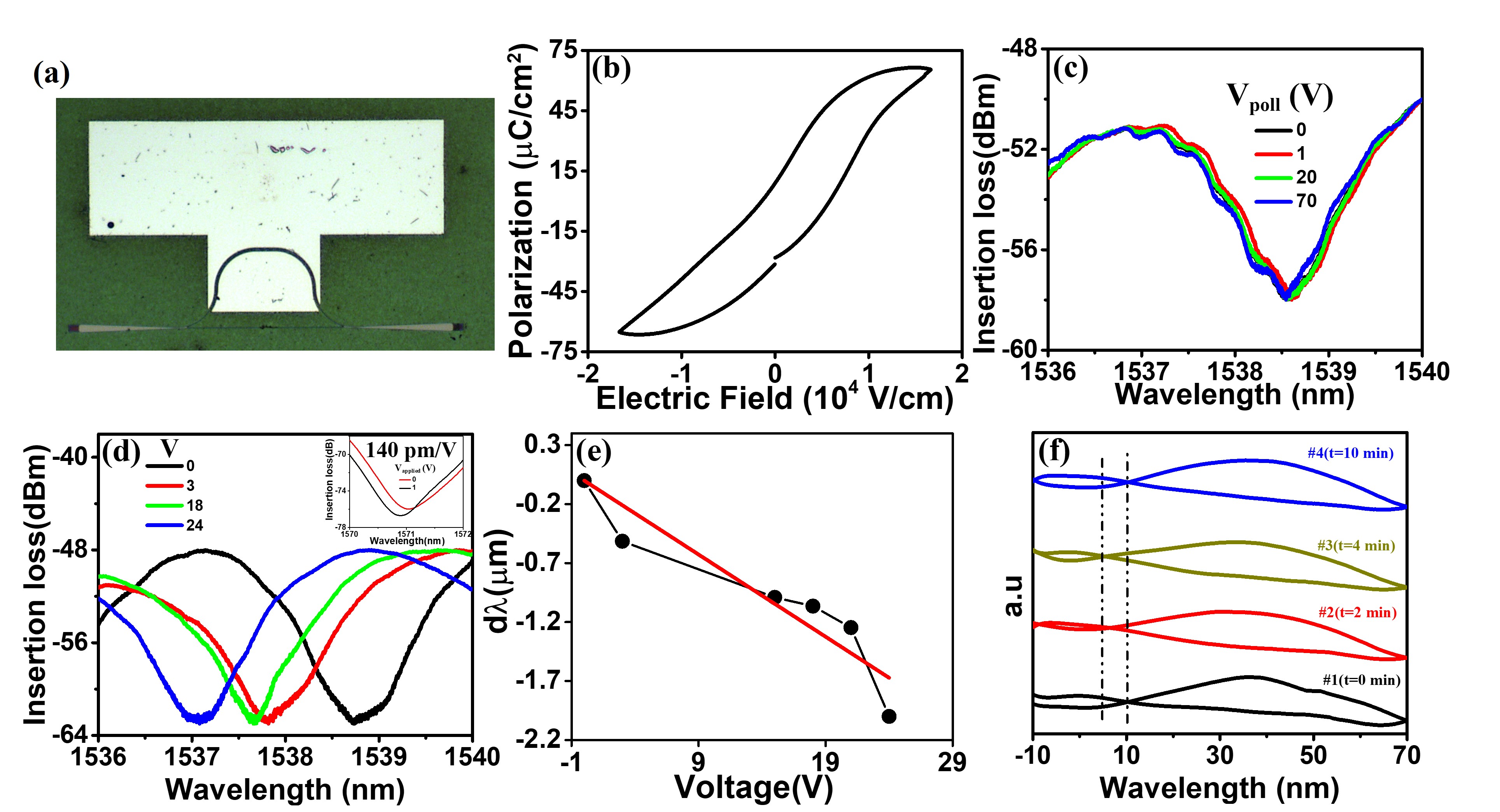}
\caption{(a) Schematic showing conventional PZT-based EO modulator design and proposed design with optical microscopy image of the fabricated MZI; (b) P-E loop of the fabricated MZI device; EO measurement of the device (c) pre-poling,(d) post-poling; (e) curve showing a blue shift in the optical spectra on the application of applied voltage and (f) C-V curve of the fabricated device showing the time stability of the ferroelectric domain on poling.}
\label{fig:EO}
\end{figure}
AFM characterization is shown in Fig.\ref{fig:device_char}(a) shows obtained "t" of 110 nm with a $\Lambda$ of 774 nm and waveguide width and height of 1.3 $\mu$m and 485 nm respectively with a roughness of \textbf{XX}. The dimensions obtained were simulated giving a per coupler efficiency of -10 dB/coupler which is $\approx$7.8 dB less than the simulated one and  a 6dB bandwidth of 40 nm which closely matches with the simulated design. The difference is attributed to the shallow etch waveguide as well as the grating height that is 73\% off from the optimized value of 411 nm as seen in Fig.\ref{fig:device_char}(b). Fig.\ref{fig:Passive}(a) shows the optical characterization of waveguide with varying lengths with Fig.\ref{fig:Passive}(b) giving us a loss of -0.056 dB/nm. The losses in the waveguide can be attributed to the usage of doped amorphous Si as the waveguide. Fig.\ref{fig:Passive}(c \& d) shows the optical characterization of the fabricated ring and MZI device with the ring radius of 250 nm with 0 nm ring to waveguide gap (merged during fabrication process) and unbalanced MZI with a $\Delta$L of 180 nm.

\subsection{Electro-optic device characterization}
Fig.\ref{fig:EO}(a) shows the optical microscopic image of the fabricated MZI interferometer. The conventional approach uses PZT as active material on Si with an intermediate buffer layer while the current proposed architecture for EO modulator uses PZT as the platform itself to build the photonic device. The advantage of the current approach over the conventional approach is that the optical field interacts with the electrical field without an intermediate buffer layer and hence increasing the electro-optic interaction. Fig.\ref{fig:EO}(a) shows an optical image of the fabricated Si MZI-based EO modulator on PZT platform. Fig.\ref{fig:EO}(b) shows a PE loop hysteresis with a peak polarization of 60 $\mu$C/cm$^2$ with a coercive field of $\approx$10 kV/cm confirming the piezo-electric nature of the PZT film post-fabrication process. Fig.\ref{fig:EO}(c) and (d) show the EO response of the fabricated MZI pre and post-poling. The poling process involved applying a fixed bias of 70 V on the EO probes for 1 Hr and allowing them to cool for 45 min before starting measurement. The pre-poled device shows a negligible shift in the optical spectra while we see a blue shift in the spectra post-poling evident from Fig.\ref{fig:EO}(c) and (d) respectively. The slope of the shift on the application of voltage is plotted in Fig.\ref{fig:EO}(e) giving a stabilized shift of $\approx$71 pm/V with peak shift being 140 pm/V shown in the inset of Fig.\ref{fig:EO}(d). The reduction of a shift in the spectrum is attributed to the poling stability wherein the PZT material losses its poled state during low voltage measurement as is evident from the C-V curve in Fig.\ref{fig:EO}(f). The C-V curve on consecutive high voltage sweep gets poled which is marked by the shift in the C-V curve to the left which returns to its original position on a sweep after 10 min of time proving the loss in the orientation of the ferroelectric domain in PZT.\cite{su2011poling} Table.\ref{tab:Works on PZT based EO modulator} shows the comparison of works done on a PZT-based EO modulator with the DC spectrum shift being comparable to the best-reported values on dielectric buffer layers. The efficiency of the present work can further be improved by high-temperature poling\cite{chen2022high} allowing us to have a stable domain poling at lower applied voltage as well as reducing the gap between the electrodes for poling and measurements.

\begin{table}[]
\centering
\caption{Works on PZT based EO modulator }
\label{tab:Works on PZT based EO modulator}
\begin{tabular}
{|p{2cm}|p{4cm}|p{2cm}|p{2cm}|p{4cm}|}
\hline
Platform & PZT integration and orientation &  Buffer layer & Deposition Method & DC Shift (pm/V)\\
\hline
\raisebox{-\totalheight}{Si\cite{george2015lanthanide}} & Top, (100) & Lanthanides & sol-gel & 50-200 (1.2$\mu$m - ellipsometry) \\
\hline  
\raisebox{-\totalheight}{Sapphire\cite{ban2021high}} & Top, (100) & LaNit	 & sol-gel & 80 (n$_{PZT}$= 2.4) \\
\hline 
\raisebox{-\totalheight}{Si\cite{nakada2009lanthanum}}	& Top, (100) & perovskites		 & Aerosol	  & -\\
\hline
\raisebox{-\totalheight}{Si\cite{ban2022low}} & Top, (100)& LaNit			 & Sol-gel  & 96  \\
\hline
\raisebox{-\totalheight}{Si\cite{singh2021sputter}} & Top, Poly-crystalline  & MgO & Sputter  & 14 \\
\hline
{SiN\cite{alexander2018nanophotonic}} & Top, -  & Lanthanide & Sol-gel  & 13.5 \\
\hline
PZT/MgO{$^{\textbf{*}}$} & Bottom, (100)  & None		 & Sputter  & 71 (140 maximum) \\
\hline
\end{tabular}
\end{table}

\section{Conclusion}

We have successfully demonstrated a CMOS compatible PZT platform for Si photonic device fabrication. We have optimized PZT on MgO for high orientation with a surface roughness of less than 2 nm enabling. Si waveguide was simulated with an efficiency of -2.2 dB/coupler for TE mode and -3 dB/coupler for TM mode, which is the first reporting on PZT platform. We report a tailored grating coupler that can work as a TE coupler, TM coupler or a polarization independent coupler. Fabrication was done with the Si grating efficiency of around -11 dB/coupler and a 6 dB bandwidth of 30 nm. The simulated waveguide with fabricated dimension matched well with a loss of -12 dB/coupler and bandwidth of 40 nm. Electro-optic characterization was done for MZI. The shift achieved was 71 pm/V, which is the first-time report on PZT platform. The performance of the modulator can be enhanced by improving the fabrication methodology and making the physical dimension closer to the simulated ones and reducing the gap between the electrodes of the modulator.

\section*{Acknowledgments}
SKS thanks Professor Ramakrishna Rao chair fellowship.

\section*{Disclosures}
The authors declare no conflicts of interest.

\section*{Data availability}
Data underlying the results presented in this paper are not publicly available at this time but may be obtained from the authors upon reasonable request.

\bibliographystyle{unsrt}  
\bibliography{biblio_file}

\end{document}